\documentstyle[preprint,aps]{revtex}
\begin{document}
\draft
\title{Inelastic Neutron Scattering Study of Mn$_{12}$-Acetate}
\author{Yicheng Zhong and M. P. Sarachik}
\address{Physics Department, City College of the City University of 
New York, New York, New York 10031}
\author{Jonathan R. Friedman}
\address{Department of Physics and Astronomy, The State University of New York 
at Stony Brook, Stony Brook, NY 11794}
\author{R. A. Robinson, T. M. Kelley, H. Nakotte\cite{hn} and A. C. Christianson}
\address{Manuel Lujan Jr. Neutron Science Center, Los Alamos National 
Laboratory, Los Alamos, NM 87545}
\author{F. Trouw}
\address{Intense Pulsed Neutron Source, Argonne National Laboratory, Argonne, 
IL 60439}
\author{S. M. J. Aubin and D. N. Hendrickson}
\address{Department of Chemistry and Biochemistry, University of California 
at San Diego, La Jolla, CA 92093}
\date{\today}
\maketitle
\begin{abstract}

We report zero-field inelastic neutron scattering experiments on a 
deuterated powder sample of Mn$_{12}$-Acetate consisting of a large number of 
nominally identical spin-$10$ magnetic clusters.  Our resolution enables us 
to see a series of peaks corresponding to transitions between the 
anisotropy levels within the spin-$10$ manifold.  A fit to the spin Hamiltonian 
$H = -D S_z^2-\mu_B{\bf B}\cdot {\bf g}\cdot {\bf S}-AS_z^4+C(S_+^4+S_-^4)$ yields an anisotropy 
constant $D =(0.54\pm 0.02)$ K and a fourth-order diagonal anisotropy coefficient 
$A=(1.2\pm 0.1)\times 10^{-3}$ K (the other terms being negligible).  
Performed in the absence of a magnetic field, 
our experiments do not involve the $g$-values as fitting parameters, thereby 
yielding particularly reliable values of $D$ and $A$.
\end{abstract}
\pacs{PACS numbers: 78.70.N, 29.30.H, 36.40.Cg, 75.45.+j}

\noindent {\bf INTRODUCTION}

High-spin molecular magnets provide a unique laboratory for the study of 
Quantum Tunneling of Magnetization (QTM).  To date, the most intensively 
studied system of this type is Mn$_{12}$-Acetate, 
[Mn$_{12}$O$_{12}$(CH$_3$COO)$_{16}$(H$_2$O)$_4$]$\cdot 2$CH$_3$COOH$\cdot4$H$_2$O 
(hereafter referred to as Mn$_{12}$-Ac).  
First synthesized by Lis\cite{lis}, it consists of Avogadro's number of 
weakly interacting\cite{weakint}, chemically identical Mn$_{12}$-Ac 
molecules residing on a body-centered tetragonal 
lattice.  The magnetic core of each molecule contains four Mn$^{4+}$ ($S=3/2$) 
and eight Mn$^{3+}$ ($S=2$) ions which form an $S=10$ ground state at low 
temperatures\cite{spin10}.  A strong magnetocrystalline anisotropy results 
in a double-well potential with each molecule's $(2S+1)=21$ states yielding two 
degenerate ground states $m=\pm 10$, and a set of doubly degenerate excited 
states $m=\pm 9$,$\pm 8$....(except for $m=0$)\cite{spin10,double}.  Below the 
blocking temperature of $\approx 3$ K, a remarkable series of steps were found 
in the hysteresis loops of oriented-powder samples at regular intervals of 
magnetic field, steps which were interpreted as a manifestation of 
QTM\cite{friedman}.  Experimental confirmation of these steps was 
provided soon thereafter in studies of single crystals\cite{thomas}.

Other experimental evidence supports this interpretation\cite{fominaya,luis1} 
but there is no general agreement on the mechanism responsible for the QTM in 
Mn$_{12}$-Ac.  Up to fourth-order terms, the spin Hamiltonian of the 
system can be written as:
\begin{equation} 
H= -DS_z^2-\mu_B{\bf B}\cdot {\bf g}\cdot {\bf S}-AS_z^4+C(S_+^4+S_-^4)=H_0+H'
\label{equation}
\end{equation}
where $D$ is the anisotropy constant, the second term represents the Zeeman 
energy, and the remaining are higher-order terms in the crystalline anisotropy.  
$H_0= -DS_z^2-g_{para}\mu_B B_zS_z-AS_z^4$ includes all terms that commute 
with $S_z$ and do not give rise to tunneling; 
$H'= -g_{perp}\mu_BB_xS_x+ C(S_+^4+S_-^4)$ represents symmetry-breaking 
terms that could 
give rise to tunneling, associated with a transverse magnetic field and 
transverse fourth-order anisotropy terms.  Major efforts 
are currently underway to determine the relative importance of magnetic fields 
and crystalline anisotropy in 
accounting for the relaxation rates observed in 
Mn$_{12}$-Ac\cite{garanin,luis2,fort}.  Accurate, reliable experimental 
determinations of the spin Hamiltonian, 
Eq. (1), thus provide crucial information.

EPR measurements performed recently in Mn$_{12}$-Ac have yielded two 
different sets of values for the coefficients $D$ and $A$ of Eq. (1).  Barra 
{\it et al.}\cite{barra} measured high-field EPR spectra at frequencies ranging from 
$150$ to $525$ GHz in magnetic fields up to $25$ T on a polycrystalline powder 
sample, yielding $g_\parallel =(1.93\pm 0.01)$, $g_\perp=(1.96\pm 0.01)$, 
$D =(0.56\pm 0.04)$ K and $A=(1.1\pm 0.1)\times 10^{-3}$ K.  Using 
high-sensitivity EPR techniques in the frequency range between $35$ and $115$ 
GHz, Hill {\it et al.}\cite{hill} 
studied a submillimeter single crystal; their results imply 
$D =0.59$ K and $A=0.88\times10^{-3}$ K\cite{hillcalc} with $g_\parallel$ 
ranging from 1.97 to 2.08 and $g_\perp=1.9$.

EPR measurements are normally done in a magnetic field and the $g$-values, 
generally unknown, are treated as (additional) fitting parameters.  
In contrast, neutron scattering experiments are normally 
performed in the absence of external magnetic fields, and yield a more 
direct determination of the coefficients $A$ 
and $D$.  An inelastic neutron 
scattering study by Hennion {\it et al.}\cite{hennion} of 
partially deuterated Mn$_{12}$-Ac found   
a well-defined peak around $0.3$ THz ($1.24$ meV) which was attributed to 
excitations from $m=\pm 10$ to $m=\pm 9$.  The peak broadens 
on its low energy side as the temperature increases, but these 
authors were unable to resolve any detailed structure.

In the present study, we have performed zero-field inelastic neutron 
scattering  experiments on fully deuterated Mn$_{12}$-Ac.  The excitation 
spectra were measured 
with relatively uniform and high resolution at finite neutron 
energy transfer up to $20$ meV, covering the excitation energies 
of the spin-10 manifold of Mn$_{12}$-Ac. Since $g$-factors do not 
enter the problem in the absence of a magnetic field, this method 
allows a more accurate determination of the spin Hamiltonian.

\noindent {\bf EXPERIMENTS AND RESULTS}

A 14-gram deuterated Mn$_{12}$-Ac powder sample was prepared for 
the inelastic neutron scattering experiments.  The sample was 
characterized following the method of 
ref. 5 and steps at the same values of magnetic field were 
seen in its hysteresis loops.  We used the PHAROS chopper 
spectrometer\cite{robinson} at the LANSCE spallation neutron source at Los 
Alamos, covering energy transfers between $0$ and $20$ meV with 
resolutions of $0.4$ and $0.8$ meV FWHM (full-width-at-half-maximum) 
at two different incident energies (12 and 20 meV), and temperatures between 
$1.4$ and $77$ K.  We also used 
QENS\cite{bradley}, an inverse geometry crystal analyzer spectrometer at 
the Intense Pulsed Neutron Source at Argonne National Laboratory with final 
neutron energy of $3.63$ meV at five different temperatures ranging from 
$1.4$ K up to $30$ K.  Its energy resolution is $\approx 100 \mu$eV FWHM.

Data taken at temperatures of 1.4 K, 10 K, 17 K and 30 K are shown in Fig. 1.  
The large maximum centered about zero energy is due to elastic scattering.  At 
$1.4$ K, a single sharp peak is observed at 
$1.24$ meV; we attribute this to excitations from  spin 
states $m=\pm 10$ to $m=\pm 9$.  We note that at $1.4$ K, the overwhelming 
majority of spins 
are in the ground states $m=\pm 10$.  As the temperature is raised and some 
of the spins are thermally activated to higher energy states, new peaks 
develop on the low energy side of the $1.24$ meV peak; we attribute these to 
transitions from $m=\pm 9$ to $\pm 8$, $\pm 8$ to $\pm 7$, etc.  Transitions 
such as those between $m=\pm 9$ and $\pm 7$ are forbidden by neutron scattering 
selection rules, $\Delta S=0,\pm 1, \Delta m=0,\pm 1$.  Due to the increased 
population of higher energy levels at higher 
temperature, peaks also appear that are symmetrically placed with respect to $E=0$ 
on the neutron energy-gain side.  No maxima appear above $1.24$ meV up 
to $\approx 3$ meV, where further excitations occur that are possibly associated 
with transitions between different spin manifolds\cite{hennion}; this 
confirms that the peak at $1.24$ meV corresponds to transitions between the 
ground and first excited states of the spin-10 manifold.  The maxima are 
labelled by the index $m$, which denotes the level from which 
each excitation occurs; thus, the $1.24$ meV peak is 
labeled $10$, the adjacent peak, $9$, and so on.  As shown below, the positions 
of these peaks contain key information regarding 
the spin Hamiltonian of Mn$_{12}$-Ac.

Since there is no externally applied magnetic field in our 
experiments, and the Zeeman energy due to the internal magnetic 
field of Mn$_{12}$-Ac (estimated to be several hundred Oe\cite{hartmann}) 
is at least two orders of magnitude smaller than the 
anisotropy energy, the term $-\mu_B{\bf B}\cdot {\bf g}\cdot {\bf S}$ in spin Hamiltonian (1) can 
be safely neglected.   Furthermore, the fourth-order transverse 
anisotropy term $C(S_+^4+S_-^4)$ has little effect on the Eigen-energies 
of the states with large $|m|$. The energy of the states 
probed in our experiments near the bottom of the anisotropy wells can 
thus be approximated by $E_m=-Dm^2-Am^4$, and the energy of 
excitation from levels $\pm m$ to $\pm (m-1)$ will be
\begin{equation}
\Delta E_m = E_{m-1} - E_m = D(2m-1) + A [m^4-(m-1)^4].
\label{equation}
\end{equation}  
In Fig. 2, six excitation energies are plotted as a function 
of the index $m$.  The deviation from linear dependence clearly 
indicates the importance of including a diagonal fourth-order 
term.  A two-parameter fit to Eq. (2) gives 
$D=(4.67\pm 0.18)\times 10^{-2}$ meV=$(0.54\pm 0.02)$ K and 
$A=(1.04\pm 0.10)\times 10^{-4}$ meV=$(1.2\pm 0.1)\times 10^{-3}$ K.  These values 
are very close to the EPR 
results obtained by Barra {\it et al.}: $D=(0.56\pm 0.04)$ K, 
$A=(1.1\pm 0.1)\times 10^{-3}$ K.

\noindent {\bf DISCUSSION}

For our values of $D$ and $A$, 
the full height of the anisotropy barrier 
(defined as the energy difference between $m=0$ and $m=10$) is 
calculated to be $(66\pm 3)$ K.  The ratio $|Am^4/Dm^2|\approx 0.2$ for $m=10$.  Since 
$A$ and $D$ have the same sign, the level spacings near the bottom 
of the anisotropy wells are relatively sparser,and the distribution of 
levels near the top of the barrier denser.  Due to the presence of a 
fourth-order term, the energy  levels 
will not come into resonance simutaneously for a given field 
applied along the anisotropy axis.  Since two levels of 
different quantum numbers $m$ and $m'$ are degenerate  when 
$H=-(m+m')[D+A(m^2+ m'^2)]/(g_{para}\mu_B)$, all pairs of states $m'= -m$ 
come into resonance simultaneously only in zero magnetic field.  
Moreover, the spacing between steps in the hysteresis loops will not 
be constant.  Detailed comparison of the magnetic fields at which 
maxima in the relaxation rate occur with the calculated level crossings using the 
parameters given above can, in principle, allow a determination of the 
specific levels near the top of the barrier that participate in the tunneling.

In summary, we have used zero-field inelastic neutron scattering 
to probe the excitation spectrum of Mn$_{12}$-Ac.  Our resolution enables us to 
observe a series of peaks within the range from $0$ to $1.24$ meV; 
we attribute these peaks to transitions within the $S=10$ manifold.  
A two-parameter fit yields 
values for the anisotropy constant $D$ and the coefficient of the fourth-order 
diagonal anisotropy $A$ that are inconsistent with those deduced from EPR 
experiments of Hill {\it et al.} and agree well with results of Barra 
{\it et al.}\cite{barra}.

Work was supported at City College by NSF grant DMR-9704309 and at the 
University of California, San Diego by NSF grant DMR-9729339.  Work at Los 
Alamos and Argonne National Laboratories was supported by the U.S. 
Department of Energy, Office of Basic Energy
Sciences, under contracts W-7405-ENG-36 and W-31-109-ENG-38, respectively.

\begin{figure}
\caption{Neutron scattered intensity versus energy at temperatures of 
1.4, 10, 17 and 30 K, taken on the QENS spectrometer at Argonne National Laboratory.}
\label{fig1}
\end{figure}
\begin{figure}
\caption{Energies of the peaks of Fig. 1 plotted as a function of index $m$ 
($m$ denotes initial states for energy loss and final states for energy gain).}
\label{fig2}
\end{figure}
\end{document}